\documentclass{article}
\usepackage[T1]{fontenc} 
\usepackage[utf8]{inputenc} 
\usepackage{ismir,amsmath,cite,url}
\usepackage{graphicx}
\usepackage{color}

\usepackage{booktabs}
\usepackage{wasysym} 
\usepackage{makecell} 

\title{Text-to-Music Models for Chiptune Video Game Music Generation}

\multauthor
{Igor Mattos dos Santos Varejão\textsuperscript{\ensuremath{\sharp}\eighthnote} \hspace{0.3cm} 
Felipe Ferreira Marra\textsuperscript{\ensuremath{\flat}\eighthnote} \hspace{0.3cm} 
João Carlos Santana Silveira\textsuperscript{\ensuremath{\sharp}}}
{ \bfseries{Lucas Nascimento Ferreira\textsuperscript{\ensuremath{\sharp}}} \\
\textsuperscript{\ensuremath{\sharp}} Departamento de Ciência da Computação, UFMG -- Belo Horizonte, MG, Brazil \\
\textsuperscript{\ensuremath{\flat}} Departamento de Informática, UFV -- Viçosa, MG, Brazil\\
\textsuperscript{\eighthnote}{\normalsize Shared first authorship}\\
{\tt\small \{igor.varejao,joaocarlossantana,lferreira\}@dcc.ufmg.br, felipe.marra@ufv.br}
}

\def\authorname{I. Varejão, F. Marra, J. Silveira, L. Ferreira}

\usepackage[bookmarks=false,pdfauthor={\authorname},pdfsubject={\papersubject},hidelinks]{hyperref}

\begin{document}

\maketitle

\begin{abstract}
Modern text-to-music (T2M) models can generate music across different genres, but few have been explored for video game music. In this paper, we investigate autoregressive T2M models for Nintendo Entertainment System (NES) music, as the NES provides a standardized video game music space where all pieces are composed for four synthesizers. NES music is accessible in multiple representations: audio signal, MIDI, and synth machine code (NLM). Thus, we propose three transformer models with a shared architecture: an audio model and two symbolic models for MIDI and NLM. To support our audio model, we first evaluate EnCodec and DAC as neural audio codecs for NES music. While DAC outperforms EnCodec, both codecs perform worse on NES music than on general non-chiptune music (MTG-Jamendo). Using DAC as the neural codec for the audio model, we compare our three models against a MusicGen baseline. Results show that all of them outperform the baseline, and that the symbolic models outperform the audio-based one, even when it is pre-trained on MTG-Jamendo, suggesting that state-of-the-art neural codecs struggle to reconstruct chiptune video game music.
\end{abstract}

\section{Introduction} 
\label{sec:introduction}

Text-to-music (T2M) is the task of generating music from textual descriptions. For example, given the prompt ``a piano playing a sad melody,'' a model must generate a composition that matches this description. T2M models have been explored in both the audio \cite{music_gen_copet_simple_2024, music_lm_agostinelli_musiclm_2023} and symbolic \cite{lu2023musecocogeneratingsymbolicmusic, bhandari2024text2midigeneratingsymbolicmusic} domains, with state-of-the-art (SOTA) approaches being able to generate high-quality music across a wide range of genres. However, the question of how audio-based T2M models perform on chiptune music remains open. Investigating chiptune generation is relevant because, while the genre emerged during the 8-bit era of video games, it remains highly popular today, particularly within indie video game soundtracks \cite{reid2025chiptune}. Furthermore, from an audio modeling perspective, chiptune music is composed entirely of synthetic instruments. These instruments produce distinct waveforms that represent an out-of-distribution challenge for models trained primarily on acoustic or non-chiptune datasets, making them difficult to reproduce.

In this paper, we investigate chiptune T2M models using the Nintendo Entertainment System Music Database (NES-MDB), a dataset that contains 5,278 music pieces from the soundtracks of 397 NES games, with each piece represented in three formats: audio, MIDI, and synth machine code \cite{donahue2018nes}. Within the NES-MDB context, the NES synth machine code is called the NES Language Modeling (NLM) format. 
Since NES-MDB does not have textual descriptions associated with the music pieces, we generate textual descriptions with a large language model (LLM) from extracted music features and game metadata. Thus, our dataset is composed of 5,278 examples in the format $(w_i, m_i, n_i, d_i)$, where $w_i$, $m_i$, and $n_i$ are the audio waveform, MIDI representation, and NLM representation of example $i$, and $d_i$ is its textual description.

We propose three similar encoder-decoder transformer architectures: one for the MIDI format, one for the NLM format, and one for the audio domain. All models use a pre-trained T5 encoder to encode textual music descriptions into semantic embeddings. The symbolic ones (MIDI and NLM) decode NES music similarly to text models by predicting the next vocabulary token. The audio model decodes music into quantized audio tokens produced by a neural codec model, following MusicGen's codebook interleaving strategy \cite{music_gen_copet_simple_2024}.

For our experiments, we first compare the average reconstruction error of two SOTA neural codecs, Descript Audio Codec (DAC) \cite{dac_kumar_high-fidelity_nodate} and EnCodec \cite{encodec_defossez_high_2022}, to find which performs better on NES-MDB. Since NES-MDB contains only chiptune music, we also compare these two neural codecs on MTG-Jamendo \cite{mtg_jamendo_bogdanov2019mtg}, a popular dataset of non-chiptune music containing 55,525 tracks from 87 different genres. Results show that DAC outperforms EnCodec on both NES-MDB and MTG-Jamendo. However, when analyzing the impact of the datasets, both neural audio codecs perform better on MTG-Jamendo than on NES-MDB.

Using the DAC codec for the audio model, we train our three models on our annotated NES-MDB dataset and compare them against a MusicGen baseline. For the audio model, we consider two versions, one trained solely on NES-MDB and another pre-trained on MTG-Jamendo and later fine-tuned on NES-MDB. We evaluate all models with the Fréchet Audio Distance (FAD) for audio quality, the Kullback--Leibler divergence (KLD) for composition quality, and the CLAP score for how well the music matches the text description. All of our models beat the baseline on every metric, and the symbolic MIDI model scores best on all three. Pre-training helps the audio model, improving on the from-scratch version on every metric. However, it still trails the symbolic approaches on text alignment and falls far behind them on audio quality. This gap is partly expected, since symbolic outputs are rendered by a synthesizer that does not add noise. We hypothesize that current SOTA neural codecs struggle to compress chiptune music, making it hard for generative models to produce realistic music directly as an audio waveform. The data, code and prompts to reproduce our results are available on: \url{https://github.com/ivarejao/nes-music-gen.git}

\section{Related Work} \label{sec:related_works}

\subsection{The NES Music Database} \label{sec:nesmdb_dataset}

NES-MDB is a collection of chiptune music specifically sourced to facilitate machine learning research in music generation and analysis \cite{donahue2018nes}. The dataset comprises 5,278 distinct musical pieces extracted directly from the original soundtracks of 397 NES games. A key feature of NES-MDB is that it provides a highly standardized acoustic and symbolic space. This standardization stems from the strict hardware constraints of the original NES Audio Processing Unit (APU), which limits polyphony to four primary synthetic sound channels: two pulse wave generators (typically used for melody and harmony), one triangle wave generator (primarily used for basslines), and one pseudo-random noise generator (used for percussive elements).

To support different modeling approaches, NES-MDB provides exact alignments of its musical pieces across three distinct representations: audio waveforms, MIDI, and NLM, a specialized sequence format consisting of a timestamped list of instructions that control the synthesizer state machine.

\subsection{Text-to-Music Models} \label{sec:t2m_models}

The T2M task aims to generate music pieces that semantically align with a given textual description. In the audio domain, SOTA architectures typically leverage neural audio codecs, such as SoundStream \cite{soundstream_zeghidour_soundstream_2022}, EnCodec \cite{encodec_defossez_high_2022}, and the Improved RVQ-GAN (DAC) \cite{dac_kumar_high-fidelity_nodate}, to compress long audio sequences into shorter token sequences, which are then modeled autoregressively. For instance, MusicLM \cite{music_lm_agostinelli_musiclm_2023} leverages the shared text-music embedding space of MuLan \cite{huang2022mulanjointembeddingmusic} to condition its generation process. Following a similar autoregressive paradigm, MusicGen \cite{music_gen_copet_simple_2024} predicts multiple codebook tokens in parallel to improve computational efficiency, utilizing a pre-trained T5 model \cite{t5} as its textual encoder. Conversely, Stable Audio \cite{evans2025stable} adopts a diffusion-based architecture, utilizing CLAP \cite{elizalde2023clap} embeddings to condition the generation. Notably, while models like MusicLM and MusicGen are designed to process unconstrained natural language descriptions, Stable Audio primarily relies on structured musical tags for its textual conditioning.

In the symbolic domain, approaches frame music generation as a sequence modeling task over discrete musical events. For instance, MuseCoco \cite{lu2023musecocogeneratingsymbolicmusic} generates symbolic music from text descriptions using a two-stage framework consisting of text-to-attribute understanding and attribute-to-music generation. More recent architectures utilize end-to-end training on text-annotated MIDI datasets, such as MidiCaps \cite{midicaps}, to streamline this pipeline. The Text2midi architecture \cite{bhandari2024text2midigeneratingsymbolicmusic}, for example, utilizes an encoder-decoder setup where a pre-trained FLAN-T5 \cite{chung2022scalinginstructionfinetunedlanguagemodels} encoder processes captions to condition an autoregressive transformer decoder that generates REMI+ \cite{remiplusvonruette2023figaro} encoded MIDI tokens. Alternatively, the MIDI-LLM architecture \cite{midillm_Huang} adopts a decoder-only approach by expanding the token embedding layer of a pre-trained LLM to directly incorporate MIDI tokens. This vocabulary expansion preserves the original model's parameter structure, enabling the use of existing inference-time optimizations.

\section{Text-to-Music Chiptune Generation} \label{sec:t2m_chiptune_gen}

This section details the dataset annotation process and the base architecture used for our three T2M models, as well as the different classification heads employed to model the various music formats.

\subsection{Annotating the NES Music Database}

Our annotation pipeline consists of two main stages. First, we extract musical features and game metadata from each NES-MDB piece. We then incorporate this information into a prompt for an open-weight LLM, which generates a fluent, natural-language description intended to resemble a human-written input. We extract 5 musical features (key, time signature, tempo, duration, and instruments) and 3 game attributes (game title, game genre, and console).

In the first stage, following the MidiCaps methodology \cite{midicaps}, we extract key and time signature from MIDI using Music21 \cite{music21}, compute tempo and duration using Mido \cite{mido_python}, and retrieve instruments directly from MIDI files. To increase descriptive diversity, the detected tempo in BPM for each sample is randomly quantized using one of two mapping schemes: nine classical terms (from `Grave' to `Prestissimo') based on the thresholds $\{40, 60, 70, 90, 110, 140, 160, 210\}$, or four broader terms (`Slow', `Moderate tempo', `Fast', `Very fast') using the thresholds $\{80, 120, 160\}$. Similarly, duration captions are assigned by binning the track length against the thresholds $\{30, 120, 300\}$, categorizing each sample as a `Short fragment', `Short song', `Song', or `Long piece'. For the game attributes, we extract game titles and console names from NES-MDB metadata and assign game genres using the 40-category classification introduced in \cite{nesmvdb_cardoso_nes_2024}. The music features enable music-based prompting (e.g., ``Generate a fast piece with a catchy melody on a pulse wave synth''), while the game attributes support game-based prompting (e.g., ``Generate a piece for a platforming game in the style of \textit{Super Mario Bros.}'').

In the second stage, we inject the music features and game attributes of each piece into a carefully designed prompt for a locally deployed \textsc{mistral-7B-instruct-v0.1} LLM to generate a textual description of the input piece. To produce more natural descriptions, we provide four captions of chiptune songs from the MidiCaps dataset \cite{midicaps} as few-shot examples. Furthermore, we emphasize in the prompt that the generated description should only include the information given as input, without adding any external information. Lastly, we instruct the LLM to vary the sentence structure across pieces to ensure more diversity in the descriptions. As an example of a description in our dataset, the track ``Cantina'' from \textit{Star Wars: The Empire Strikes Back} is described as follows:

\begin{quote}
\itshape
This is a short fragment of music from the game ``StarWars: The Empire Strikes Back'' on the NES console. The game genre is an action platformer. The music is composed of pulse waves 1 and 2, a triangle wave, and noise percussive. The key is D minor. The tempo is moderate, at 120 beats per minute. The time signature is 4/4. The duration of the piece is 13 seconds.
\end{quote}

\subsection{Chiptune Models}

All our models follow a similar encoder-decoder transformer architecture, with a pre-trained T5 encoder~\cite{t5} to encode the input textual descriptions of the music. The decoder of each model is a traditional transformer decoder~\cite{vaswani2017attention}, with cross-attention applied to the T5 text embeddings. What changes is how each model represents the music data. For the MIDI model, the music data is tokenized via the Multi-Track representation scheme from MIDI-GPT~\cite{pasquier2025midi}, which has a vocabulary size of 565 tokens. For the NLM model, on the other hand, the music data is tokenized via byte-pair encoding (BPE), configured to yield a vocabulary of the same size as the MIDI model (565 tokens). For the audio model, we follow MusicGen's approach~\cite{music_gen_copet_simple_2024}; however, instead of EnCodec~\cite{encodec_defossez_high_2022} as the audio tokenizer, we use DAC, since it outperforms EnCodec in the NES domain (see Table~\ref{tab:codecs_recons_loss}). The DAC codec has a vocabulary size of 1027 audio tokens.

The decoders of the MIDI and NLM models are exactly the same, with a softmax classification head over their respective symbolic vocabularies. For the audio model, we use a decoder that predicts only the first 4 of the 9 DAC codebooks, following a delay pattern~\cite{music_gen_copet_simple_2024}. Each codebook has its own embedding table, but these embeddings are summed, allowing the transformer decoder to work without the codebook dimension. The output of the audio transformer decoder is fed to 4 linear classification heads, each the size of a DAC codebook, allowing the model to recover the codebook dimension.

\section{Experiments and Results} \label{sec:exps}

We first evaluate the reconstruction error of EnCodec and DAC to decide which neural codec to consider for our audio model. We then evaluate our three encoder-decoder models (audio, MIDI, and NLM) with respect to objective metrics of audio quality, composition quality, and semantic alignment with the input textual descriptions. For the audio model, we consider two versions, one trained solely on NES-MDB and another first pre-trained on MTG-Jamendo and then fine-tuned on NES-MDB.

\subsection{Experimental Setup}
The text encoder is a pre-trained T5-Base~\cite{t5}, with 12 transformer layers, 12 attention heads, and a hidden dimension of 768. Our transformer decoders have 12 layers, 16 attention heads, and a hidden dimension of 1024. We train all models for at most 30 epochs, using early stopping with a patience of 5 and a batch size of 8. T5 remains frozen during the training of all models.
For the audio model, MTG-Jamendo pre-training is limited to 70 epochs, and all decoder weights are updated during fine-tuning.

We train all models on an NVIDIA A100 GPU with 80~GB of memory. Our optimizer is AdamW, with a learning rate of $5\times 10^{-4}$, betas of $0.9$ and $0.95$, and weight decay of $0.1$. We also leverage a cosine scheduler with a linear warmup of 3000 steps (approx. one epoch). We use a cross-entropy loss and remove the text conditioning for 30\% of the batches. We keep the same dataset split as NES-MDB: training ($85.3\%$), validation ($7.7\%$), and test ($7\%$). However, we merge the training and validation sets to increase the amount of training data.

For inference, each model has its own temperature, classifier-free guidance (CFG), and top-k hyperparameters, selected via grid search, as shown in Table~\ref{tab:grid_search}. This search minimizes the mean of the min--max normalized CLAP (with inverted sign), FAD, and KLD scores, computed over 32 samples for each hyperparameter configuration.

\begin{table}[t]
 \centering
\begin{tabular}{lccc}
\toprule
& \textbf{Temperature} & \textbf{CFG} & \textbf{Top-K} \\
\midrule
Audio Scratch & $1.5$ & $1.0$ & $50$ \\
Audio Pre & $1.5$ & $2.5$ & $100$ \\
MIDI & $1.5$ & $1.5$ & $200$ \\
NLM & $1.5$ & $1.5$ & $100$ \\
\bottomrule
\end{tabular}
\caption{Hyperparameters calculated via grid search for each model.}
\label{tab:grid_search}
\end{table}

\subsection{Dataset Augmentation}
To expand the dataset and provide broader context for our models, we augment the original MIDI samples before generating the textual descriptions. We modify two parameters: tempo ($T$) and key ($K$). Specifically, we increase or decrease the tempo by 20\% and transpose the key up or down by one semitone. Given the possible values $T \in \{0.8, 1.0, 1.2\}$ and $K \in \{-1, 0, 1\}$, this yields a total of 9 combinations per track, including the original. After generating these variations in the MIDI domain, we use the NES-MDB Python package \cite{donahue2018nes} to synthesize the corresponding NLM and audio representations. This augmentation is applied only to the training set; thus, the dataset expands from 5,278 to 38,928 samples.

\subsection{Codec Evaluation}

We follow DAC's evaluation methodology to compare DAC and EnCodec, measuring the reconstruction fidelity of the pieces in the test set with four metrics: Mel-spectrogram and multi-resolution STFT distances (spectral fidelity, lower is better), ViSQOL (a perceptual-quality estimate correlated with human mean opinion scores (MOS), higher is better), and scale-invariant SDR (waveform-level fidelity in dB, higher is better) \cite{dac_kumar_high-fidelity_nodate}. The comparison spans two domains: MTG-Jamendo, representing general music similar to what these codecs were originally trained on, and NES-MDB, consisting of 8-bit chiptune game audio, to probe how well each codec generalizes to a narrower, more synthetic domain relevant to the game-soundtrack setting. Results are reported as mean $\pm$ standard deviation over the respective test sets.

\begin{table}[t]
\centering
\begin{tabular}{lcc}
\toprule
 & $\text{DAC}_{4|44.1}$ & $\text{EnCodec}_{4|32}$ \\
\midrule
\multicolumn{3}{c}{\textbf{MTG-Jamendo}} \\
\midrule
Mel $\downarrow$ & $2.97 \pm 0.29$ & $5.83 \pm 0.93$ \\
STFT $\downarrow$ & $0.93 \pm 0.11$ & $1.49 \pm 0.20$ \\
ViSQOL $\uparrow$ & $4.24 \pm 0.26$ & $3.58 \pm 0.43$ \\
SI-SDR $\uparrow$ & $11.47 \pm 4.01$ & $5.66 \pm 3.79$ \\
\midrule
\multicolumn{3}{c}{\textbf{NES-MDB}} \\
\midrule
Mel $\downarrow$ & $8.13 \pm 3.45$ & $13.42 \pm 5.53$ \\
STFT $\downarrow$ & $1.38 \pm 0.28$ & $2.18 \pm 0.18$ \\
ViSQOL $\uparrow$ & $3.25 \pm 0.49$ & $2.68 \pm 0.47$ \\
SI-SDR $\uparrow$ & $9.48 \pm 3.70$ & $8.05 \pm 3.67$ \\
\bottomrule
\end{tabular}
\caption{Reconstruction quality of neural audio codecs on MTG-Jamendo and NES-MDB. Values are mean $\pm$ standard deviation. Subscripts denote the number of codebooks and the sample rate in kHz, respectively.}
\label{tab:codecs_recons_loss}
\end{table}

Table~\ref{tab:codecs_recons_loss} shows that $\text{DAC}_{4|44.1}$ consistently outperforms $\text{EnCodec}_{4|32}$ on both datasets and across all four metrics, using the same number of codebooks. On MTG-Jamendo, DAC achieves lower Mel and STFT distances (2.97 vs. 5.83 and 0.93 vs. 1.49, respectively) and higher ViSQOL and SI-SDR scores (4.24 vs. 3.58 and 11.47 vs. 5.66), indicating both stronger perceptual quality and waveform-level fidelity when reconstructing general, non-chiptune music. DAC's advantage over EnCodec remains on NES-MDB. However, on NES-MDB, both codecs degrade substantially relative to their MTG-Jamendo performance, suggesting that chiptune audio lies outside the distribution on which these codecs were originally trained.

\subsection{Generative Models Evaluation}

Using DAC as our neural codec, we evaluate the generative models across the three proposed music representations (audio, MIDI, and NLM) against the MusicGen (MGen) baseline, using standard metrics from the audio-based T2M literature \cite{music_gen_copet_simple_2024}: KLD (with PaSST \cite{koutini2021efficient}) for composition quality, FAD (with VGGish \cite{hershey2017cnn}) for audio quality, and the CLAP score for semantic text alignment. For each model, we generate a synthetic counterpart for three pieces from each of the 40 game genres in the NES-MDB test set, yielding 120 generated pieces per model, which are then compared against the corresponding ground-truth pieces from the test set.

As shown in Table~\ref{tab:results_generation}, our models outperform MGen across all three metrics, with CLAP increasing from $-0.03$ to values between $0.27$ and $0.46$. The MIDI model achieves the best results on every metric, with the lowest KLD ($0.44$), the lowest FAD ($5.89$), and the highest CLAP score ($0.46$). NLM ranks second in both FAD ($9.16$) and CLAP ($0.40$), but its KLD ($0.84$) is higher than that of Audio Pre ($0.65$), which obtains the second-best KLD. Comparing the two audio models, we observe that pre-training on MTG-Jamendo improves all three metrics relative to training from scratch. However, the FAD of both audio models remains roughly an order of magnitude higher than that of the symbolic models. Overall, these results suggest that, while pre-training improves audio models, they still lag behind symbolic ones, highlighting an important research gap.

\begin{table}[t!]
\centering
\begin{tabular}{lccc}
\toprule
 & KLD $\downarrow$ & FAD $\downarrow$ & CLAP $\uparrow$ \\
\midrule
MGen & $1.08$ & $95.61$ & $-0.03$ \\
\midrule
Audio Scratch & $0.90$ & $82.44$ & $0.27$ \\
Audio Pre & $0.65$ & $78.21$ & $0.33$ \\
MIDI & $\mathbf{0.44}$ & $\mathbf{5.89}$ & $\mathbf{0.46}$ \\
NLM & $0.84$ & $9.16$ & $0.40$ \\
\bottomrule
\end{tabular}
\caption{Performance evaluation across multiple music representations compared to the MusicGen baseline. Bold values indicate the best performance.}
\label{tab:results_generation}
\end{table}

\section{Conclusion and Future Work} 
\label{sec:conclusion}

We investigated T2M generation for video game soundtracks using the NES-MDB dataset. We compared DAC and EnCodec as neural audio codecs, finding that DAC yields better reconstruction fidelity. Using DAC as the audio tokenizer, we trained and compared text-conditioned generative models across audio, MIDI, and NLM representations, showing that symbolic representations outperform audio-based ones, even when the audio model is pre-trained on a large dataset. We hypothesize that this gap is partly due to the difficulty current neural audio codecs have in faithfully modeling chiptune music. For future work, we will extend our evaluation with subjective metrics and additional codecs, and explore fine-tuning existing codecs on chiptune audio.

\bibliography{ISMIRtemplate}

\end{document}